\documentclass[prl]{revtex4}
\usepackage{graphicx}
\usepackage{bm}
\usepackage{color}
\usepackage{bm}
\definecolor{pink}{rgb}{1,1,0} 
\definecolor{yellow}{rgb}{1,1,0}
\definecolor{orange}{rgb}{1,0.5,0}
\definecolor{white}{rgb}{1,1,1}

\begin{document}
\title{Can we hear Kolmogorov spectra? \footnote{Borrowed to Alan C. Newell.} }
\author{ Gustavo  D\"uring$^1$, Christophe Josserand$^2$ and  Sergio Rica$^{1,3}$}
\affiliation{$^1$ Departamento de F\'\i sica, Universidad de Chile, Blanco Encalada 2008, Santiago, Chile.\\
$^2$Laboratoire de Mod\'elisation en M\'ecanique,
CNRS UMR 7607, Case 162, 4 place Jussieu, 75005 Paris, France\\
$^3$ LPS--ENS,  CNRS UMR 8550, 24 Rue Lhomond, 75231 Paris Cedex 05, France.}
\begin{abstract}
We study the long-time evolution of waves of a thin elastic plate in the limit of small deformation so that modes of oscillations interact weakly. According to the theory of weak turbulence a nonlinear wave system evolves in long-time creating a slow transfer of energy from one mode to another. We derive this kinetic equation for the spectral transfer in terms of the second order moment. We show explicitly that such a non-equilibrium theory describes the approach to an equilibrium wave spectrum, and describes also an energy cascade, often called the Kolmogorov-Zakharov spectrum. We perform numerical simulations confirming this scenario. 
\end{abstract}

\maketitle  
\medskip

{\it Introduction.--} Since more than forty years it was established that long time statistical properties of a random fluctuating wavy system  possess  a natural asymptotic closure because of  the dispersive nature of waves and the weakly nonlinear wave interaction \cite{hasselmann,benney}.  Indeed this so-called ``weak turbulence theory" has shown to be a powerful method to study the evolution of nonlinear dispersive wave systems. 
It results that the longtime dynamics is driven by a kinetic equation for the distribution of spectral densities. This method, was developed for surface gravity waves \cite{hasselmann,zakgrav66}, surface capillary waves \cite{zakcap67}, plasma waves \cite{zakplasma67}, nonlinear optics \cite{Dyachenko-92}, {\it etc.}

The actual kinetic equation has non-equilibrium properties similar to the usual Boltzmann equation for dilute gases, thus it conserves energy, momentum, and exhibits an H-theorem driving the system to equilibrium, characterized by the named Rayleigh-Jeans distribution. Most important, besides the elementary equilibrium (or thermodynamic) solution, Zakharov has shown \cite{zakplasma67} that power law non-equilibrium solutions also arise, namely the Kolmogorov--Zakharov (KZ) solutions or KZ spectra, which describe the exchange of conserved quantities ({\it e.g.} energy) between large and small length scales.

Experimental evidence of KZ spectra have been found in ocean surface  \cite{atmos} and in  capillary surfaces waves \cite{putterman,danish, brazhnikov}. On the other hand, numerical simulations of surface waves shown the realization of  KZ  spectrum for weak turbulent capillary waves \cite{zakcapnum} and, more recently, for gravity waves \cite{zakgravnum}.

In this article  an oscillating thin elastic plate or shell is considered.  Adding inertia to the well known (static)  theory of thin plates one finds ballistic dispersive waves \cite{landau}, which interact via nonlinear terms that are weak if the plate deformations are small.  A previous work has dealt with solitary wave propagation on the surface of a cilindrical shell \cite{putterman2} due to the balance between dispersive bending waves and nonlinearities. 
We develop thus a weak turbulence theory for the surface deflection. Shortly,  
the bending waves travel randomly over the system and interact resonantly between each other {\it via} 
the weak nonlinearities. The mathematics beyond the resonant condition are formally identical to the conservation of energy and momentum in classical gas of particles. In this sense an elastic plate with is formally equivalent to a 2D gas of classical particle interacting with a non-trivial scattering cross-section.  Indeed, an isolated system evolves from a random initial condition to a situation of statistical equilibrium like a gas of classical particles does.
As usually, in addition to statistical equilibrium for isolated systems, the weak
 turbulence theory predicts here an energy cascade from a source of energy (a driving forcing) to a 
 dissipation scale typically characterized by plastic deformations. 
 Moreover, while there is often a lack of direct observations of weak turbulence predictions, we exhibit
 numerically these behaviors for the plate dynamics.

\begin{figure}[hc]
\begin{center}
\centerline{\includegraphics[width=4cm]{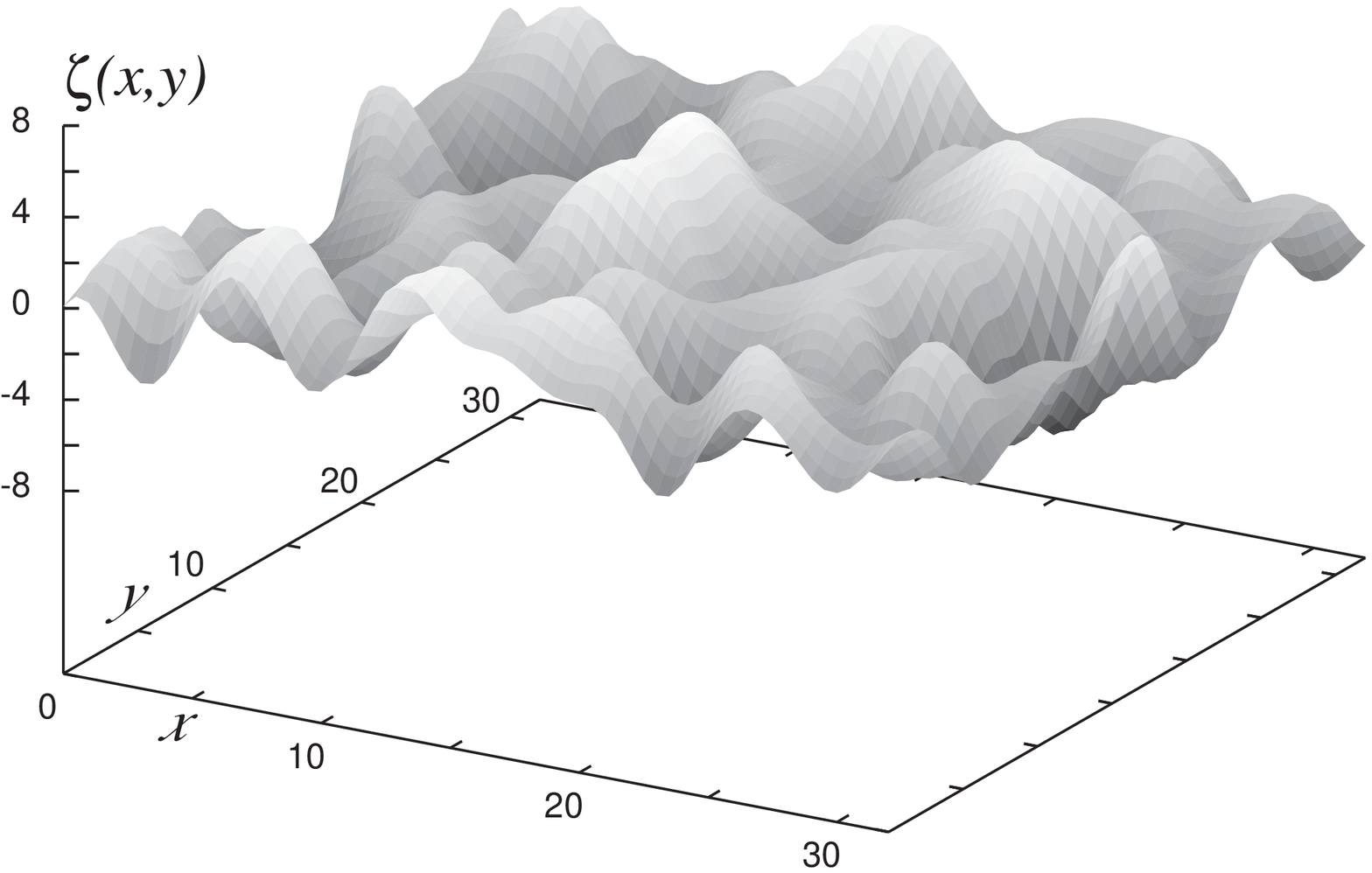} \quad\includegraphics[width=4cm]{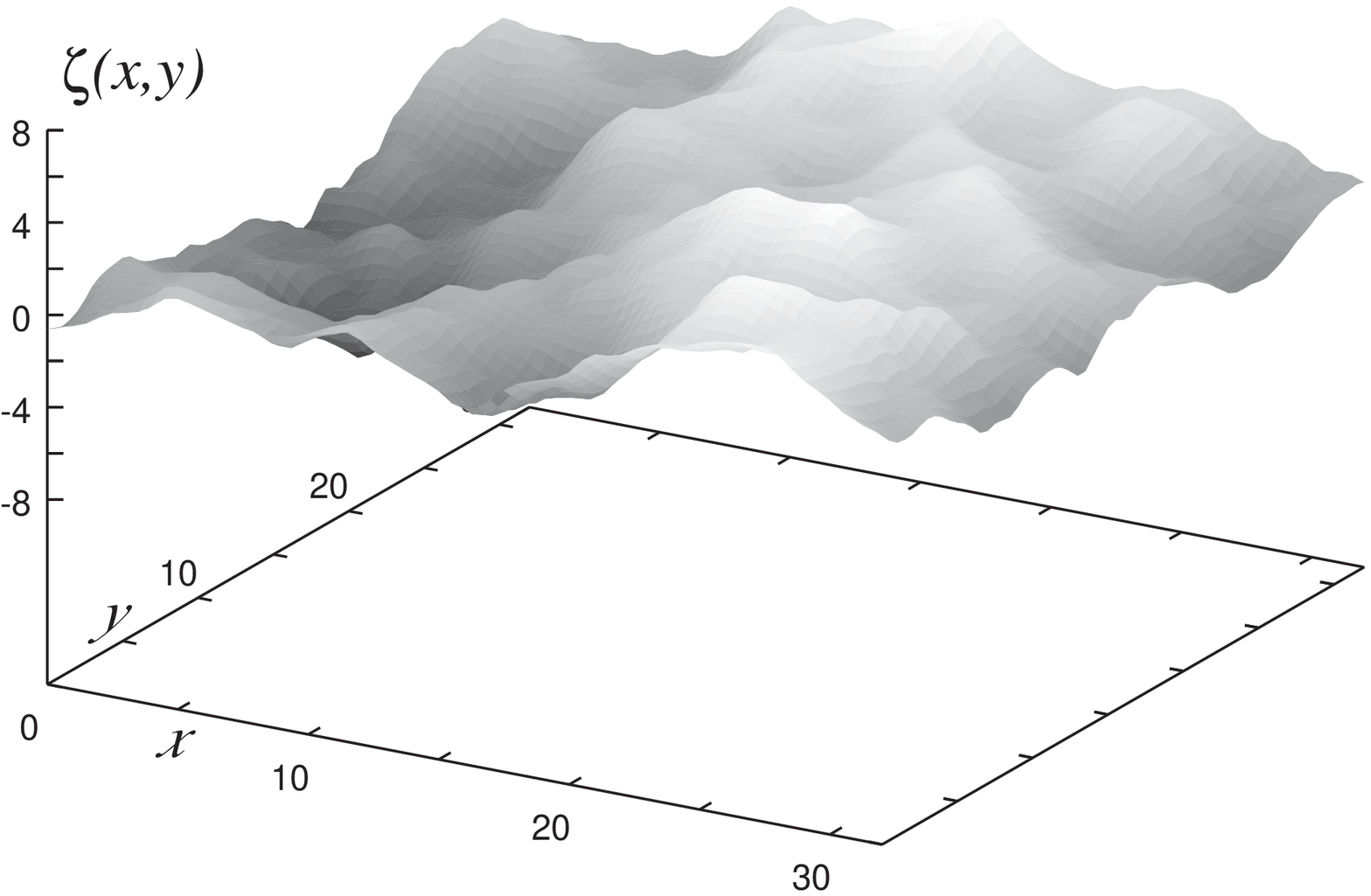} }
\caption{ \label{fig1} 
Zoom over a portion of the surface plate  deflection $\zeta(x,y)$, the left had image is the initial condition while the right hand image represents a late evolution of the elastic plate. The aspect ratio is 1:1:1.
}
\end{center}
\end{figure}

This dynamics is illustrated in Fig. \ref{fig1} for an isolated dissipation free system where the plate deformation are shown at initial time and after long evolution. 

{\it Theory.--}
The starting point is the dynamical version of the F\"oppl--von K\'arm\'an equations
\cite{fvk,landau} for the amplitude of the deformation $\zeta(x,y,t)$ and for  the Airy stress function $\chi(x,y,t)$:
\begin{eqnarray} 
\rho\frac{\partial^2 \zeta}{\partial t^2} &=& - \frac{Eh^2}{12(1-\sigma^2)}\Delta^2\zeta +
\{\zeta,\chi\}  ;
\label{foppl0}\\
\frac{1}{E}\Delta^2\chi &=&- \frac{1}{2}\{\zeta,\zeta\}.
\label{foppl1}
\end{eqnarray}
Where  $h$ is the thickness of the elastic sheet. The material has a  mass density $\rho$, a Young modulus $E$. Its Poisson
ratio is $\sigma$.
$\Delta=\partial_{xx}+\partial_{yy}$ is the usual Laplacian and the bracket $\{\cdot,\cdot\}$ is defined by
$\{f,g\}\equiv f_{xx}g_{yy}+f_{yy}g_{xx}-2f_{xy}g_{xy},$  which is an exact divergence, so equation (\ref{foppl0}) preserves the momentum of the center of mass, namely $\partial_{tt} \int \zeta(x,y,t) dx dy =0$. Equation (\ref{foppl1}) for  the Airy stress function
$\chi(x,y,t)$ may be seen as the compatibility equation for the in--plane stress tensor which follows the dynamics.

Small plane waves perturbations ($\zeta \sim e^{i( {\bm k}\cdot {\bm x} -\omega t)}$ with ${\bm x} =(x,y)$) of a plane plate are dispersive with the usual ballistic behavior of bending
waves: $
\omega_{\bm k} = hc |\bm k|^2= hck^2.
$
Where $c=\sqrt\frac{E}{12(1-\sigma^2)\rho}$ has dimension of a speed.

{\it Weak turbulence theory.-} Despite the complexity of (\ref{foppl0}) and (\ref{foppl1}) the system 
presents a hamiltonian structure which is straightforward in Fourier space. Defining the Fourier transforms as $\zeta_{\bm k}(t)= \frac{1}{2\pi} \int \zeta({\bm x},t) e^{i  {\bm k}\cdot {\bm x} } d^2{\bm x}$ (with $\zeta_{\bm k}= \zeta_{-{\bm k}}^*$), then one gets  from (\ref{foppl1}):
$\chi_{\bm k}(t)=  -\frac{E}{2 |{\bm k}|^4 } \{\zeta,\zeta\}_{\bm k}$
where $\{\zeta,\zeta\}_{\bm k}$ is the Fourier transform of $\{\zeta,\zeta \} $. The dynamics then reads:

\begin{eqnarray} 
\rho\frac{\partial^2 \zeta_{\bm k}}{\partial t^2} &=&  - \frac{ E h^2k^4}{12 (1-\sigma^2)} \zeta_{\bm k} \nonumber \\
 &+&    \int V_{- \bm k,\bm k_2;\bm k_3,\bm k_4}    \zeta_{{\bm k}_2}   \zeta_{{\bm k}_3}  \zeta_{{\bm k}_4} \delta^{(2)}({\bm k}-{\bm k}_2-{\bm k}_3-{\bm k}_4) d^2{\bm k}_{2,3,4} \nonumber
\end{eqnarray}
where $ d^2{\bm k}_{2,3,4}\equiv d^2{\bm  k}_2d^2{\bm  k}_3d^2{\bm k}_4$ and
$
V_{12;34}=\frac{E}{2(2\pi)^{2}} \frac{ |{\bm k}_1\times{\bm k}_2|^2  |{\bm k}_3\times{\bm k}_4|^2} {|{\bm k}_1+{\bm k}_2|^4} $.
The hamiltonian structure becomes evident, if we define as canonical variables the deformation $\zeta_{\bm k}(t)$ and the momentum $p_{\bm k}(t)= \rho \partial_t \zeta_{\bm k}(t)$.

The canonical transformation $ \zeta_{\bm k} =  \frac{X_k}{\sqrt{2}} ( A_{\bm k}+ A^*_{-{\bm k}} )$ and
$p_{\bm k} = -\frac{i}{\sqrt{2}X_k} ( A_{\bm k} - A^*_{-{\bm k}} )$ with $X_{k}  =  \frac{1}{\sqrt{\omega_ k  \rho}} $ allows to write the wave equation in a diagonalized form: $\frac{d A_{\bm k}}{d t}+i\omega_{ k} A_{\bm k}= i N_3( A_{\bm k})$ where $N_3(\cdot)$ is the cubic nonlinear interaction term.


This nonlinear oscillator has two distinct scales of time, the rapid oscillation 
$i\omega_{ k} A_{\bm k}$ and the weak nonlinearity. If the gradient of the deformation is small,
 the nonlinear terms come up to next order.  Then, after the change $A_{\bm k} = a_{\bm k}(t) e^{-i \omega_k t}$ which removes the rapid linear oscillating term, we obtain:
\begin{equation}
\frac{d a^s_{\bm k}}{d t}=-is \sum_{s_1 s_2 s_3} \int J_{-{\bm k} {\bm k}_1{\bm k}_2{\bm k}_3} e^{i t(s\omega_k-s_1 \omega_{k_1}-s_2 \omega_{k_2}-s_3 \omega_{k_3} ) }  {a^{s_1}_1 a^{s_2}_2 a^{s_3}_3 } \delta^{(2)}({\bm k}_1+{\bm k}_2+{\bm k}_3-{\bm k}) d^{2}{\bm k}_{123}
\label{ecuacionesdehamilton3}
\end{equation}
where we define $a^s_{\bm k}$ with the two possible choices $ s=+,-$ corresponding to propagation direction, such that  $ a^+_{\bm k}\equiv a_{\bm k}$ while $ a^-_{\bm k}\equiv a^*_{-{\bm k}}$.  The interaction term reads: $J_{k_1,k_2;k_3,k_4} = \frac{1}{6} X_{k_1}X_{k_2}X_{k_3}X_{k_4} {\mathcal{P}}_{234}V_{{\bm k}_1,{\bm k}_2;{\bm k}_3,{\bm k}_4}$ with ${\mathcal{P}}_{234}$ the six possible permutations between 2, 3 \& 4. Next step consists  to write a hierarchy of equations for the averaged moments: $\left< a^{s_1}_{ \bm k_1} a^{s_2}_{{\bm k}_2} \right> $, $\left< a^{s_1}_{ \bm k_1} a^{s_2}_{{\bm k}_2} a^{s_3}_{ \bm k_3} a^{s_4}_{{\bm k}_4}\right> $, {\it etc\ae tera}:
$ \frac{d}{d t} \left< a^{l_1}_{ \bm p_1} a^{l_2}_{{\bm p}_2} \right> = i  \epsilon^2{\tilde N}_3\left(  \left<a^{s_1}_{\bm k_1} a^{s_2}_{\bm k_2} a^{s_3}_{\bm k_3} a^{s_4}_{\bm p_4} \right> \right)$, {\it etc\ae tera}. A multi-scale analysis provides a natural asymptotic --over long times-- closure, for higher moments: the fast oscillations drive the system close to Gaussian statistics and higher moments are written in terms of the second order moment:
$\left< a_{ \bm k_1} a^{*}_{{\bm k}_2} \right> = n_{ {\bm k}_1} \delta^{(2)}(k_1+k_2)$, where $n_{ \bm k}$ is called the wave spectrum. 

The wave-spectrum satisfies thus a Boltzmann-type kinetic equation showing the exchange of energy from one mode to another in longtime due to the four waves resonance: 
 \begin{eqnarray}
\frac{dn_{ {\bm p}_1}}{dt}&=& 12\pi sgn(t)  \int |J_{{\bm p}_1{\bf k}_1{\bm k}_2{\bm k}_3}|^2  \sum_{s_1s_2s_3} n_{{\bm k}_1}n_{{\bm k}_2}n_{{\bm k}_3}n_{{\bm p}_1}\nonumber\\
&\times&\left(\frac{1}{n_{{\bm p}_1}}+\frac{s_1}{n_{{\bm k}_1}}+\frac{s_2}{n_{{\bm k}_2}}+\frac{s_3}{n_{{\bm k}_3}}\right)\nonumber\\ 
&\times& \delta(\omega_{ p_1}+s_1\omega_{ k_1}+s_2\omega_{k_2}+s_3\omega_{k_3})  \,  \delta({\bm p}_1 + s_1{\bm k}_1+ s_2{\bm k}_2+s_3{\bm k}_3)  \, d^{2}{\bm k}_{123}
\label{ecuacioncinetica1}
\end{eqnarray}

As for the usual Boltzmann equation, Eq.(\ref{ecuacioncinetica1}) conserves ``formally'' \footnote{``Formally'' means here that the proof requires convergence of any simple integral to the exchange of integration order by the FubiniÕs theorem \cite{newell}.} 
 the total momentum per unit area ${\bm P} =  h \int {\bm k} n_{{\bm k}}(t)  \, d^2k $ and
the kinetic energy per unit area ${\cal E} = h \int \omega_k n_{{\bm k}}(t)  \, d^2k $ and exhibits a 
$H$-theorem: let be $ \mathcal S(t)=\int  \ln(n_{{\bm k}}) \, d^Dk$ the non-equilibrium entropy, then $d \mathcal S/dt \ge 0$, for increasing time\footnote{For $t<0$ one has that $d \mathcal S/dt \le 0$, driving the system to an equilibrium too, however it should be noticed that kinetic equation (\ref{ecuacioncinetica1}) is not well posed in the sense that a positive initial condition for $n_{{\bm k}}(t=0)> 0 $  could become negative.}. 
However, despite the four wave interaction type kinetic equation (\ref{ecuacioncinetica1}), the ``wave action''  ${\cal N} = \int n_{{\bm k}}(t) d^2k $ is not conserved.
The kinetic equation (\ref{ecuacioncinetica1}) describes thus an irreversible evolution 
of the wave-spectrum towards the Rayleigh-Jeans {\it equilibrium} distribution which 
reads, when ${\bm P} ={\bm 0}$: 
\begin{equation}
n^{eq}_{\bm k}= \frac{T}{\omega_k },
\label{equil}
\end{equation}
where $T$ is called, by analogy with thermodynamics, the 
temperature (with units of energy/lenght, that is a force)  which is naturally related to the initial energy by 
${\cal E}_0 = h  \int \omega _k n_{eq} d^2\bm k= h T \int  d^2\bm k$. The quantity $ \int  d^2\bm k$ is the number of degrees of freedom per unit surface. Therefore each degree of freedom takes the same energy: $hT$. Naturally, for an infinite system this number diverges (as well as the energy). This classical
 Rayleigh-Jeans catastrophe is always suppressed due to some physical cut-off which corresponds here
 to dissipation processes for wave-lengths smaller than $h$. Numerical simulations 
 on regular grid provide also a natural cut-off $k_c=\pi/dx$ with $dx$ the mesh size, which gives 
 ${\cal E}_0 = \pi  h T k_c^2$ for a large plate.
 
{\it Kolmogorov Spectra.-} In addition, isotropic non-equilibrium distribution solution can also arise \cite{zakplasma67}. They have a major importance in the non equilibrium process for the energy sharing among scales. Those solutions can be guessed via a dimensional analysis argument but, as it was shown by Zakharov \cite{zakplasma67}, they are exact solutions of the kinetic equation. 
 Indeed, integrating over the angles the scattering amplitude $|J_{{\bm k}_1{\bm k}_2{\bm k}_3{\bm k}_4}|^2 \delta^{(2)}({\bm k}_1+{\bm k}_2+{\bm k}_3+{\bm k}_4)$ a new scattering amplitude depends only on the moduli $k_i=|{\bm k}_i|$:

$$S_{k_1,k_2,k_3,k_4} = 
= 
\int \frac{|J_{{\bm k}_1{\bm k}_2{\bm k}_3{\bm k}_4}|^2 }{| {\bm k}_2\times{\bm k}_3|}d\varphi_4. $$
 This integral can be performed providing a resonance condition involving the angle between $ {\bm k}_2 \;\& \;{\bm k}_3 $ and $\varphi_4$ via $k_2^2 + k_3^2+ 2 {\bm k}_2 \cdot {\bm k}_3 = k_1^2 + k_4^3+ 2k_1 k_4 \cos\varphi_4$. An explicit expression for $S$ is computed and since the degree of homogeneity of $|J|^2$ in $k$ is zero, $S$ scales like $1/k^2$.

Looking for a power law solution of the form $n_k = A k^{-\alpha}$, one has, that the eight terms of the collisional integral in the r.h.s. of equation (\ref{ecuacioncinetica1}) decomposes into ${\cal C}oll_{2+2} + {\cal C}oll_{3+1} $, where~:
\begin{eqnarray}
{\cal C}oll_{2+2} & =& 36 \pi A^3 \int_{\Omega_{up}}  k_2dk_2  k_3 d k_3 S_{kk_1k_2k_3}   \nonumber\\
&\times&k_1^{-\alpha}k_2^{-\alpha}k_3^{-\alpha}k^{-\alpha}\left( k^{\alpha}+k_1^{\alpha}-k_2^{\alpha}-k_3^{\alpha}\right )\nonumber\\ 
&\times&\left(1+(k_1/k)^{3\alpha-4}-(k_2/k)^{3\alpha-4}-(k_3/k)^{3\alpha-4}\right ) \nonumber\\ 
{\cal C}oll_{3+1} & =& 12\pi A^3 \int_{\Omega_{down}} k_2dk_2  k_3 d k_3 S_{kk_1k_2k_3}   \nonumber\\
&\times&k_1^{-\alpha}k_2^{-\alpha}k_3^{-\alpha}k^{-\alpha}\left( k^{\alpha}-k_1^{\alpha}-k_2^{\alpha}-k_3^{\alpha}\right )\nonumber\\ 
&\times&\left(1-(k_1/k)^{3\alpha-4}-(k_2/k)^{3\alpha-4}-(k_3/k)^{3\alpha-4}\right ).  \nonumber
\end{eqnarray}
In ${\cal C}oll_{2+2}$  the integration domain is over $\Omega_{up} = \{ 0\leq k_2 \leq k \, \& \,\sqrt{k^2-k_2^2} \leq k_3\leq k \} $ and  $k_1^2 = k_2^2 + k_3^2 -k^2$ while in ${\cal C}oll_{3+1}$ the integration is over ${\Omega_{down}} = \{ 0\leq k_2 \leq k \, \& \, 0\leq k_3\leq  \sqrt{k^2-k_2^2} \}$, with $k_1^2 = k^2 -k_2^2 -k_3^2$.

The collisional terms scaling follow  ${\cal C}oll_{2+2} = C_1(\alpha) k ^{2-3 \alpha}$ and ${\cal C}oll_{3+1} = C_2(\alpha)k^{2-3 \alpha} $. The coefficients $C_{1/2}(\alpha)$ are pure real functions depending only on $\alpha$. Both coefficient vanish with double degeneracy  at $\alpha=2$ indicating that the Kolmogorov spectrum coincides with the Rayleigh-Jeans solution (\ref{equil}): 
$n^{KZ}_{\bm k}\sim  \frac{1}{k^2 } .$
In fact,
this degeneracy reveals the existence of a logarithmic corrections, similarly to the nonlinear Schr\"odinger equation in 2D \cite{Dyachenko-92}, thus: 
\begin{equation}
n^{KZ}_{ k}= C^{te}  \frac{ h P^{1/3 } \rho^{2/3}}{(12(1-\sigma))^{2/3}}   \frac{\ln^z(k)}{k^2}.
\label{KZ}
\end{equation}
Here $P$ is the energy flux imposed in the energy cascade between the longwave scales and the short-ones (it has dimensions of mass/time$^3$), $C$ and $z$ are pure real numbers.

\begin{figure}[hc]
\begin{center}
\includegraphics[width=8cm]{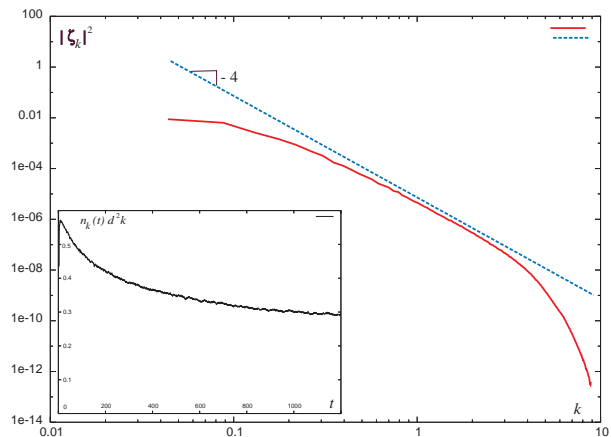}
\caption{ \label{fig2} 
Numerical simulation for a square plate with $h/L = 10^{-3}$ and using $1024^2$ modes. We plot the power spectrum of mean deflection  $ \left< |\zeta_k|^2 \right>$ versus wave number $k$ after 1200 time units. The the line plots the Rayleigh-Jeans power law $1/k^4$. The inset plots the magnitude of the violation of wave action conservation.
}
\end{center}
\end{figure}

For $\alpha=0$ and $3\alpha-4=0$ the collisional part ${\cal C}oll_{2+2} $ also vanishes. Those solutions are  the wave action equipartition ($\alpha=0$) and a second KZ spectrum $n_k \sim 1/k^{4/3}$ corresponding to
wave action inverse cascade. However, those spectrum do not vanish the second part of the collision term ${\cal C}oll_{3+1} $, in agreement with the non conservation of the wave action mentioned above. Therefore, an important consequence  is the non existence here of this second inverse cascade $n_k\sim 1/k^{4/3}$, as  usually found for four wave interaction systems such as gravity waves or the nonlinear Schr\"odinger equation.
Nevertheless, we have observed that for elastic plates the wave action conservation is only weakly violated as seen in direct numerical simulations (inset of Fig. \ref{fig2}).

{\it Numerics.--} We are first performing numerical simulations of the full non linear system of PDE (\ref{foppl0}) and (\ref{foppl1}) with no forcing nor dissipation terms.
 We have taken $c=1$, and the ratio $h/L$, $L$ being the size of a square plate, is the only dimensionless parameter in the numerics. We consider a regular grid with periodic boundary conditions. Taking advantage of the structure of the equation in Fourier space a pseudo-spectral scheme is
implemented using FFT routines \cite{fftw}, namely:
$\ddot \zeta_{\bm k}= - \omega_k^2  \zeta_{\bm k} +  \{ \zeta,\chi \} _{\bm k}$.
The linear part of the dynamics is calculated exactly: 
 $\zeta_{\bm k}(t+\Delta t) = \zeta_{\bm k}(t)\cos(\omega_k \Delta t) + \frac{ \dot \zeta_{\bm k}(t)}{\omega_k} \sin(\omega_k \Delta t).$
 The nonlinear terms in (\ref{foppl0}) and (\ref{foppl1}) are first computed in real space and
the integration in time is then performed in Fourier space with an Adams-Bashford scheme, which interpolates the nonlinear term of (\ref{foppl0}) as a polynomial function of time (of order one in the 
present calculations). Energy is conserved as best as 1/100.
For initial conditions we have taken:  $\zeta_k = \zeta_0 e^{-k^2/k_0^2} e^{i\varphi_k}$ with $\varphi_k$ a random phase, and a zero velocity field $\dot\zeta_k =0$.
 As time evolves, the random waves oscillates with a disorganized behavior, as shown in FIG. \ref{fig1}.
After a long time evolution the system build an equilibrium distribution in agreement with the Rayleigh-Jeans $n_k\sim T/k^2$ spectrum which correspond for the plate deflection to: $ \left< |\zeta_k|^2 \right>  = X_k^2 n_k = \frac{ n_k}{\rho \omega_k} = \frac{T}{\rho h^2c^2 k^4}$ as shown in Fig. \ref{fig2}.
  
\begin{figure}[hc]
\begin{center}
\includegraphics[width=8cm]{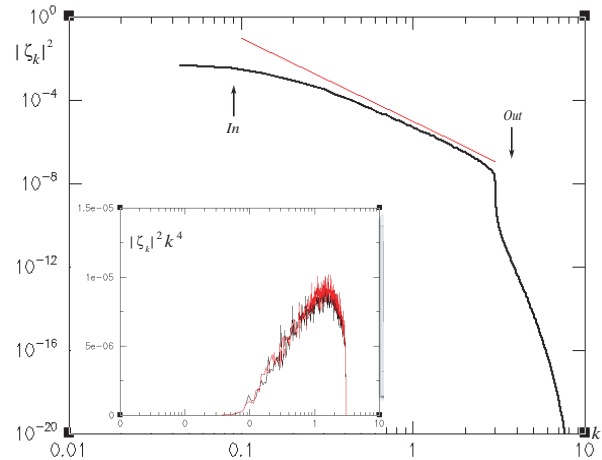}
\caption{ \label{fig3} 
Average power spectrum  $ \left< |\zeta_k|^2 \right>$ for the energy cascade,  the line plots the power law $1/k^4$. Inset plots $ k^4 \left< |\zeta_k|^2 \right> $ vs. $\log k$ showing a clear logarithmic deviation.
}
\end{center}
\end{figure}

Non equilibrium distribution may also be easily observed numerically. One requires to input energy at low wavenumbers ($k<k_{in}$) and dissipate it at large wavenumbers ($k>k_{out}$) establishing a non- equilibrium Kolmogorov spectrum at the window of transparency  $k_{in}< k<k_{out}$.  This artifact could be easily implemented, adding a $  F_{\bm k} - \gamma_k \dot \zeta_{\bm k}$ to the plate equations.  Here the forcing $  F_{\bm k}$ is a  nonzero random force for the long waves scales, and $\gamma_k$ is a fictitious linear damping,  as in \cite{zakgravnum}. The KZ spectrum is then still observed (see Fig. 
\ref{fig3}) but it presents a deviation from the $1/k^4$ spectrum in agreement with a logarithmic correction (inset of Fig. \ref{fig3}).

{\it Conclusions.--} 
We have presented a new wave system where the weak turbulence method is successfully applied.
Numerical simulations exhibit both the convergence towards statistical equilibrium for free systems
and the energy cascades when forcing and dissipation are introduced.

As amplitude of the deformations become larger, the elastic plate equations still remain valid, but 
stretching cannot be longer treated as weak perturbation and a ``wave breaking'' phenomena is expected: energy focuses into localized structures as ridges \cite{witten} and conical surfaces (named 
d-cones)\cite{pomeau}. Amazingly, a regime dominated by ridges shows a 
 power spectrum  $ |\zeta_k |^2  \sim 1/k^4  $ similar to the weak turbulence spectrum derived here. On the other hand for d-cones dominated regimes, as seemingly observed in \cite{maha}, the expected 
 spectrum should follow $ |\zeta_k |^2  \sim 1/k^6 $.  
Finally, experimental efforts seem to display a KZ spectrum close to the one predicted by weak turbulence theory \cite{hamm}.

In conclusion, since vibrating plates transmit sound  to air, we argue that a Kolmogorov spectrum can be heard! 

A. Boudaoud, E. Hamm and L. Mahadevan are acknowledged for communication their experimental results prior to publication. This work was supported by FONDECYT 1020359 and 7050143.


\begin{thebibliography}{99}
\bibitem{hasselmann}  K. Hasselmann, J. Fluid Mech. {\bf 12}  481 (1962);  {\it Ibid.} {\bf 15} 273 (1963).
\bibitem{benney} D.J. Benney, P.G. Saffman, 
Proc. Roy. Soc. London {\bf A 289},  301 (1966). 
\bibitem{zakgrav66} V.E. Zakharov, N.N. Filonenko, 
Dokl. Akad. Nauk SSSR { \bf 170}, 1292 (1966).
\bibitem{zakcap67} V.E. Zakharov, N.N. Filonenko, 
Zh. Prikl. Mekh. I Tekn. Fiz. {\bf 5}, 62 (1967)  
\bibitem{zakplasma67}V.E. Zakharov, 
Zh. Eksper. Teoret. Fiz. {\bf 51}, 686 (1966) 
\bibitem{Dyachenko-92} S. Dyachenko, 
{\it et al.,}, Physica D  {\bf 57}, 96 (1992).
\bibitem{atmos}ÊP.A. Hwang, 
ÊJ. of Phys. Oceanography  {\bf 30}, 2753 (2000).
\bibitem{putterman} W. Wright, R. Budakian, and S. Putterman, Phys. Rev. Lett {\bf 76}, 4528 (1996).
\bibitem{danish} E. Henry, P. Alstr\o m and M. T. Levinsen, Europhys. Lett., {\bf 52},  27 (2000).
\bibitem{brazhnikov} 
 M. Yu. Brazhnikov, G. V. Kolmakov, A. A. Levchenko and L. P. Mezhov-Deglin, Europhys. Lett., {\bf 58}, 510 (2002).
\bibitem{zakcapnum} A. N. Pushkarev and V. E. Zakharov, Phys. Rev. Lett. {\bf 76}, 3320
(1996).
\bibitem{zakgravnum} A. I. Dyachenko, A. O. Korotkevich and V. E. Zakharov,  Phys. Rev. Lett. {\bf 92}, 134501(2004).
\bibitem{landau} L.D. Landau and E.M. Lifshitz, Theory of Elasticity,
(Pergamon Press, New York 1959).
\bibitem{fvk} A. F\"oppl, Vorlesungen \"uber 
technische Mechanik, Bd. 5, Leipzig, 1907, p. 132. and Th. von K\'arm\'an, 
Ency. d. math. Wiss., Bd. IV. 2, II, Leipzig, 1910, \S 8.
\bibitem{putterman2} J. Wu, J. Wheatley, S. Putterman, and I. Rudnick, Phys. Rev. Lett. {\bf 59}, 2744
(1987).
 \bibitem{newell} A. Newell, S. Nazarenko, L. Biven,  Physica {\bf D 152--153} 520  (2001).
\bibitem{ZakhBook} V. E. Zakharov, V. S. L'vov and G. Falkovich,
{\it Kolmogorov Spectra of Turbulence I} (Springer, Berlin, 1992).
\bibitem{fftw} FFTW: www.fftw.org.
\bibitem{witten}T.A. Witten and H. Li, Europhys. Lett. {\bf 23}, 51 (1993).
\bibitem{pomeau}M. Ben Amar and Y. Pomeau, Proc. Soc. Lond., {\bf A 453}, 729 (1997); E. Cerda and L. Mahadevan, Phys. Rev. Lett. {\bf 80}, 2358 (1998).
\bibitem{maha} L. Mahadevan, private comunication. 
\bibitem{hamm}  A. Boudaoud and  E. Hamm, private comunication. 
\end{thebibliography}
\end{document}